\documentstyle[12pt]{article}
\advance\textheight by \topskip
\begin{document}
\rightline{SU-ITP-92-30}
\rightline{UG-10/92}
\rightline{hep-th@xxx/9212030}
\rightline{\today}

\vskip 2.5 cm
\begin{center}
{\Large\bf  SUPERSYMMETRIC STRING  WAVES}\\
\vskip 1.5 cm
{\bf E. A. Bergshoeff \footnote{Bitnet: bergshoeff@hgrrug5}}\\

\vskip 0.2cm
Institute for Theoretical Physics,
University of Groningen\\ Nijenborgh 4, 9747 AG Groningen,  The
Netherlands\\

\vskip 0.5cm
{\bf R. Kallosh \footnote{On leave of absence
 from: Lebedev Physical Institute, Moscow.
\  E-mail: kallosh@physics.stanford.edu} and T. Ort\'{\i}n
\footnote{Bitnet:
tomaso@slacvm}}\\

 \vskip 0.2cm
Physics Department, Stanford University, Stanford   CA
94305, USA

\end{center}
\vskip 1.5 cm
\begin{center}
{\bf ABSTRACT}
\end{center}
\begin{quotation}

We present plane-wave-type solutions of the lowest order
superstring effective action which have unbroken space-time
supersymmetries. They describe dilaton, axion and gauge fields in  a
stringy
generalization of the  Brinkmann metric. Some conspiracy between the
metric
and the axion field is required.\\

We show that there exists a special class of these solutions, for
which
$\alpha^\prime$ stringy corrections to the effective on-shell action,
to the
equations of motion (and therefore to the solutions themselves), and
to the
supersymmetry transformations  vanish. We call these solutions
supersymmetric  string waves (SSW).

 \end{quotation}


\newpage

\section{Introduction}

In recent years an active field of research has been the search for
and study
of non-perturbative solutions to the classical equations of motion of
superstring effective field theories and the corresponding
sigma-models.
Many bosonic solutions of such theories have been discovered. Some
special
solutions turned out to have a highly non-trivial property: although
bosonic,
they have some unbroken supersymmetries. This means that when
embedded
into the right supersymmetric theory they admit Killing spinors.

Bosonic solutions with unbroken supersymmetries in theories of
quantum
gravity are very special because they have some  kind of
supersymmetric non-renormalization property. They provide an
example of a nontrivial gravitational background in which
quantum (super)gravity corrections to the effective on-shell action
and
(sometimes) even to equations of motion vanish. Until
very
recently, the only example of a background where all perturbative
quantum
gravity corrections to the effective action vanish was given by the
empty
Minkowski space. This suggests that bosonic solutions with unbroken
supersymmetries may play an important role in the development of
quantum
gravity and may help us to understand the structure of the vacuum
state in supersymmetric theories.

Arguments that
explain these non-renormalization properties were given for
supersymmetric
string solitons \cite{Da1} and for extremal black holes \cite{Ka1}.
Another particularly interesting kind of metric is provided by the so
called
{\it pp-waves}\footnote{Here {\it pp-waves} stands for plane fronted
waves with parallel rays.}.

The purpose of this paper is to  find metrics in this class which,
together with
appropriate dilaton, axion and gauge fields, provide solutions of the
lowest order
superstring effective action and have unbroken supersymmetries. Then
we will study how
this property and the solutions themselves are affected by stringy
corrections in $\alpha^{\prime}$ \cite{Be1}.

There exists an extensive literature on this subject, and we start
by introducing {\it pp-wave} space-times and reviewing some
results relevant for our purposes in sec.2.

In  sec.3 we present a general class of solutions to the lowest order
d=10 heterotic string effective action consisting of a pp-wave
metric,
dilaton, axion and Yang-MIlls fields. We also show that half of their
supersymmetries are unbroken.

In sec.4 the stringy $\alpha^\prime$ corrections will be studied.
The metric-axion conspiracy, which was necessary for classical
supersymmetry,
will be shown to be sufficient to ensure the absence of
$\alpha^\prime$
 corrections to supersymmetry transformations and to the on-shell
effective
action. However, to get no corrections to the equations of
motion one has to constrain the solutions even more by embedding
the torsionful spin connection in the gauge group. We call such exact
solutions  {\it supersymmetric string waves}
(SSW). Finally, in sec.5 we will present our conclusions.

Appendix A presents a lemma which is used in sec.4 in the proof of
the absence of
$\alpha^{\prime}$ corrections and Appendix B contains our notation
and conventions.

\section{Gravitational plane-fronted waves}

Pp-wave geometries are space-times admitting a covariantly
constant null vector field
\begin{equation}\label{null}
\nabla_{\mu} l_{\nu} = 0\ ,  \qquad l^{\nu}l_{\nu}= 0 \ .
\end{equation}
Spacetimes with this property were first discovered by Brinkmann  in
1923 \cite{Br1}.
In four dimensions the metrics of these spaces can be written in the
general form
\begin{equation}\label{pl}
ds^2 = 2 du dv + K(u, \xi , \bar \xi ) du^2 -  d\xi d\bar \xi \ ,
\end{equation}
where $u$ and $v$ are light-cone coordinates defined by
\begin{equation}
l_{\mu}=\partial_{\mu}u\ ,\hspace{3cm}l^{\mu}\partial_{\mu}v=1\ ,
\end{equation}
(thus the metric does not depend on $v$) and $\xi = x+iy$ and $ \bar
\xi = x-iy$
are complex transverse coordinates. These metrics are classified and
described in
detail in \cite{Kr1}. Different pp-wave spaces are characterized by
different choices
of the function $K$ in eq. (\ref{pl}).  For example, when $K$ is
quadratic in $\xi$ and $\bar{\xi}$,
\begin{equation}
 K(u,  \xi , \bar \xi ) = f(u) \xi^2 + \bar f(u) \bar
\xi^2 + g(u)  \xi \bar  \xi \ .
\end{equation}
they are called {\it exact plane waves}.  Plane waves (\ref{pl}) with
$K$ of the form
\begin{equation}
K(u,  \xi , \bar \xi ) = \delta(u) f( \xi,  \bar  \xi) \ .
\end{equation}
are called {\it shock waves}.
A specific example of shock waves is given by  the  Aichelburg-Sexl
geometry \cite{AS}
\begin{equation}
K(u,  \xi , \bar \xi ) = \delta(u) \ln \, ( \xi  \bar  \xi)\ .
\end{equation}
which describes the gravitational field of a point-like particle
boosted to the speed of light.

G\"{u}ven established in 1987 \cite{Gu1} that a solution to the
lowest order
superstring effective action consisting of a straightforward
generalization of the
four dimensional exact plane waves to $d=10$, dilaton, axion and
Yang-Mills
fields has  half of the $N=1, d=10$ supersymmetries unbroken and is
also a
solution of the equations of motion of the  superstring effective
action including
all the $\alpha^\prime$-corrections. We will describe G\"{u}ven's
solution in
detail below. Investigations on the supersymmetry of plane-fronted
waves
in general relativity were made even earlier. In \cite{Gi1} there is
a
reference to unpublished work of J.  Richer who found that pp-waves
are
supersymmetric. Very general classes of pp-waves were found by Tod to
be
supersymmetric in the context of $d=4, \,  N=2$ ungauged supergravity
in 1983
\cite{To1}. Different aspects of plane wave solutions in string
theory have been
investigated in the  last few years \cite{Am1}--\cite{Ho1}.

The existence of a covariantly constant null vector field has
dramatic
consequences. For instance, for the class of d-dimensional pp-waves
with metrics
of the form  \begin{equation}\label{d}
ds^2 = 2 du dv + K (u, x^i ) du^2 - dx^i d x^i \ ,
\end{equation}
where $i, j = 1,2, ..., d-2$, the Riemann curvature is \cite{Ho1}
\begin{equation}
R_{\mu\nu\rho\sigma} = - 2 l_{[\mu}( \partial_{\nu]}
\partial_{[\rho} K )
l_{\sigma]}\ .
\end{equation}
The curvature is orthogonal to $l_{\mu}$ in all its indices.
This fact is of crucial importance in establishing that all higher
order in
$\alpha^{\prime}$ terms in the equations of motion are zero due to
the vanishing of all the
possible contractions of curvature tensors.
G\" uven added dilaton $\phi(u)$,  axion
 $H_{uij} = b_{ij}(u)$ and Yang-Mills fields $F_{ui}(u)$  which were
functions of $u$
only,  to a metric of the form eq. (\ref{d}), choosing $K(u,x^i)$
quadratic in $x^i$ (exact plane waves),  and then proved the absence
of
quantum  ($\alpha^{\prime}$) corrections to the equations of motion
\cite{Gu1}.
In addition, G\"{u}ven showed that these solutions have half of the
possible supersymmetries unbroken. Moreover he proved that the
corrections to the supersymmetry equations  vanish. This shows how
special these
solutions are, especially considering that at the time when G\"{u}ven
did his research, not  much was
known about $\alpha^\prime$ corrections.

 Note that supersymmetry
played no  role in establishing this non-renormalization
theorem. Similar results have been obtained in refs. \cite{Am1},
\cite{Ve1}, \cite{Ho1}.

In arbitrary dimension $d$ the most general metrics
admitting a covariantly constant null vector (\ref{null}),
were discovered by Brinkmann in 1925 \cite{Br2}:
\begin{equation}\label{BR}
ds^2 = 2 du dv + A_{\mu} (
u, x^i ) d x^{\mu} du - g_{ij}(
u, x^i )  dx^i d x^j \ ,  \quad l^\mu A_\mu =0\ ,
\end{equation}
where $\mu, \nu = 0,1, ...,  d-1$ and $i, j = 1,2, ..., d-2$.
The supersymmetry properties of this general metric have not yet been
studied,
although they have been discussed recently in the context of string
theory
and sigma models in \cite{Ru}, \cite{Ho1}, \cite{Ts1}. Note that the
general
Brinkmann metric (\ref{BR}) in $d=10$ has 8 functions $A_i(u, x^i )$
and 28 functions $g_{ij}(u, x^i)$ more than the
metric (\ref{d}) investigated by G\"{u}ven,
where only the $uu$-component of the metric $A_u=\frac{1}{2} K$ was
present and   was quadratic in $x^i$.

The issue of corrections to the effective string equations of motion
for
general Brinkmann metrics has been studied by Horowitz \cite{Ho1}. He
argued that
if the functions $A_i$ depend on the coordinates $x^{i}$,  the
metrics of the solutions
 do acquire corrections. If, however, this
dependence is polynomial, the number of non-vanishing terms that
correct the
equations of motion is finite for these solutions. An analogous
statement was made for the corrections to the axion field: no
corrections for
$B_{\mu\nu}$ linear in $x^i$s and  a finite number of corrections for
polynomial dependence.

We will investigate a sub-class of Brinkmann metrics more general
than the one
studied by G\"{u}ven. We will consider (\ref{BR}) with flat
transverse
space, i.e. $g_{ij}(u, x^i)=\delta_{ij}$. The functions $A_\mu$ will
be arbitrary
functions of $u$ and $x^i$. First we will identify the solutions to
the zero
slope limit of the superstring effective action ($N=1, d=10$
supergravity)
coupled to Yang-Mills and those which
have unbroken supersymmetries in this limit.
 We will look for solutions with at most one half of the
supersymmetries broken. This means that we will admit only one
algebraic
constraint, which in our case will be
\begin{equation}
\gamma^{\lambda} l_{\lambda} \epsilon =\gamma^{u} \epsilon = 0\ .
\end{equation}
\section{String Plane Waves in Ten-Dimensional Einstein\- -Yang-Mills
Supergravity}

Our starting point is the bosonic part of the action of $N=1,\, d=10$
supergravity coupled to Yang-Mills
\cite{Ch1}:
\begin{equation}\label{eq:action}
S={1\over 2}\int d^{10}x\ \sqrt {-g}e^{-2\phi}\biggl (
-R+4(\partial\phi)^2-
{1\over 3}H^2-{1\over 2}\beta\, {\rm tr} F^2\biggr )\ ,
\end{equation}
where $\phi$ is the dilaton, $F_{\mu\nu}$ is the field-strength of
the Yang-Mills field
$V_\mu:\  F_{\mu\nu}=2\partial_{[\mu}V_{\nu]}+ [V_\mu,V_\nu]$, and
$H$ is the
field-strength of the axion $B_{\mu\nu}$ which includes the
Yang-Mills Chern-Simons form:

\begin{equation}
H_{\mu\nu\rho}=\partial_{[\mu}B_{\nu\rho]}- 3 \beta{\rm tr}
\biggl \{ V_{[\mu}\partial_\nu V_{\rho]}- {1\over 3}V_{[\mu}V_\nu
V_{\rho]}\biggr \}\ .
 \end{equation}
The constant $\beta$ is related to the Yang-Mills coupling constant
$g$
by $\beta=1/g^2$. Traces are taken in the adjoint representation.

The equations of motion that follow from (\ref{eq:action}) are
\begin{eqnarray}
R_{\mu\nu}-2\nabla_\mu\partial_\nu\phi +
H_{\mu\lambda\rho}H_\nu{}^{\lambda\rho}
+\beta{\rm tr}F_{\mu\lambda}F_\nu{}^{\lambda} &=& 0 \ ,
\label{eq:eqs1}\\
R-4\nabla_\mu\partial^\mu \phi+4(\partial\phi)^2 + {1\over 3}H^2
+{1\over 2}\beta{\rm tr} F^2 &=&0\label{eq:eqs2}  \ , \\
\nabla_\lambda \bigl ( e^{-2\phi}H^{\lambda\mu\nu}\bigr )
&=&0\label{eq:eqs3}  \ ,\\
 \nabla_\lambda \bigl  (e^{-2\phi}
F^{\lambda\mu}\bigr ) +e^{-2\phi}[V_\nu,F^{\nu\mu}]
 + e^{-2\phi} H^\mu{}_{\lambda\rho}F^{\lambda\rho}&=& 0  \
.\label{eq:eqs4}
\end{eqnarray}

In order to investigate the existence of unbroken supersymmetries for
our purely bosonic
solutions we only need to consider the bosonic terms in the
supersymmetry
transformation rules of the fermions.  They are given by
\begin{eqnarray}
\delta \psi_\mu &=& \bigl (\partial_\mu - {1\over 4}
  \Omega_{+\mu}^{ab}\gamma_{ab}\bigr)
\epsilon   \ ,
   \label{eq:susy1}\\
\delta\lambda &=& \bigl (\gamma^\mu\partial_\mu\phi - {1\over 6}
                               H_{\mu\nu\rho}
\gamma^{\mu\nu\rho}\bigr)\epsilon \ ,
                              \label{eq:susy2}\\
\delta\chi &=& - {1\over 4}F_{\mu\nu}\gamma^{\mu\nu}\epsilon\ ,
                         \label{eq:susy3}
\end{eqnarray}
where  the torsionful spin connections $\Omega_{\pm \mu}^{ab}$ are
defined by
\begin{equation}\label{eq:torsion}
\Omega_{\pm \mu}^{ab} = \omega_\mu{}^{ab}(e) \pm H_\mu{}^{ab}.
\end{equation}
$\omega_\mu{}^{ab}(e)$ is the torsion-free, metric-compatible spin
connection,
and so the axion field  strength $H$  gives torsion to the spin
connections $\Omega_{\pm \mu}^{ab}$.
Our Ansatz for the metric will be Brinkmann's with flat transverse
space. In
our notations\footnote{The inverse metric can be rewritten avoiding
recursiveness as
\begin{equation}
g^{\mu\nu}= \eta^{\mu\nu}-2\eta^{\rho(\mu}\eta^{\nu)\sigma}A_\rho
l_\sigma
+\eta^{\rho\sigma}A_\rho A_{\sigma}\eta^{\mu\alpha}\eta^{\nu\beta}
l_\alpha l_\beta\ .    \nonumber
\end {equation}}
\begin{equation}\label{eq:metric}
g_{\mu\nu}=\eta_{\mu\nu}+2A_{(\mu}l_{\nu)}\ , \hskip 2truecm
g^{\mu\nu} = \eta^{\mu\nu}-2A^{(\mu}l^{\nu)}-A^2l^\mu l^\nu \ ,
\end{equation}
where $A_\mu=A_\mu(u,x^i)$, and $l_\mu$ is the covariantly constant
null vector (\ref{null}).  For
later purposes, we need to reformulate our Ansatz in terms
of zehnbein fields
\begin{equation}\label{zehnbeins}
e_\mu{}^a = \delta_\mu{}^a + A_\mu l^a \ , \hskip 2truecm
e_a{}^\mu = \delta_a{}^\mu -A_a l^\mu \ .
\end{equation}
Our Ansatz for the axion and Yang-Mills fields is
\begin{equation}
B_{\mu\nu} = 3B_{[\mu}l_{\nu]}\ , \hskip 2.8truecm  V_\mu=Vl_\mu \ .
\end{equation}
Finally, we will assume that all the fields are independent of the
coordinate $v$
but depend arbitrarily on $u$ and $x^i$. We also assume that
$A_v=B_v=0$ . In
covariant notation,
$l^{\mu}\partial_{\mu}\phi =
l^{\mu}\partial_{\mu}V=l^{\mu}\partial_{\mu}A_{\nu}
= l^{\mu}\partial_{\mu}B_{\nu} = 0$ and $l^\mu A_\mu=l^\mu B_\mu=0$.
Note that the
component $B_u$ is pure gauge but $A_u$ is not.

The Christoffel symbols corresponding to (\ref{eq:metric}), and the
 spin-connection corresponding to (\ref{zehnbeins}) are given by
\begin{eqnarray}
\bigl\{ \matrix{\rho\cr\mu\nu} \bigr\} & = & l_{(\mu}{\cal
A}_{\nu)}{}^\rho +
 \partial_{(\mu}A_{\nu)}l^\rho\ ,\hspace{3cm}
{\cal A}_{\mu\nu}=\partial_\mu A_\nu - \partial_\nu
 A_\mu \ ,    \label{Ch} \\
 \omega_\mu{}^{ab} & = & - l^{[a}{\cal A}^{b]}{}_\mu +
{1\over 2}{\cal A}^{ab}l_\mu\ ,\hspace{2.8cm}   l^\mu {\cal
A}_{\mu\nu}=0  \ .
 \end{eqnarray}
Note that the spin connection only depends on  ${\cal A}_{\mu\nu}$
but
 the Christoffel symbols (\ref{Ch}) depend also on $\partial_{(\mu}
A_{\nu)}$. Then, the curvature tensor is:
\begin{equation}
R_{\mu\nu}{}^{ab}(\omega)=
-l_{[\mu}\partial_{\nu]}{\cal A}^{ab}-l^{[a}\partial^{b]}{\cal
A}_{\mu\nu}
-l_{[\mu}{\cal A}_{\nu]}{}^\alpha {\cal A}_{\alpha}{}^{[a}l^{b]}\ .
\end{equation}
We also need the curvature with respect to the torsionful
spin connection (\ref{eq:torsion}). The torsion tensor for
our Ansatz is given by
\begin{equation}
H_\mu{}^{ab} = l^{[a}{\cal B}^{b]}{}_\mu + {1\over 2}{\cal
B}^{ab}l_\mu\
,\hspace{2cm}
{\cal B}_{\mu\nu}=\partial_\mu B_\nu -\partial_\nu B_\mu\ ,
\end{equation}
since the Yang-Mills Chern-Simons form vanishes identically in this
case.
Therefore, the torsionful spin-connections are
\begin{equation}\label{eq:torspin}
\Omega_{\pm \mu}^{ab} = -l^{[a}\bigl({\cal A}\mp{\cal
B}\bigr)^{b]}{}_\mu
+ {1\over 2}\bigl ({\cal A} \pm {\cal B}\bigr )^{ab}l_\mu\ ,
\end{equation}
and for the corresponding torsionful curvature tensor we find
\begin{eqnarray}
R_{\mu\nu}{}^{ab}\bigl (\Omega_\pm) &=&
-l_{[\mu}\partial_{\nu]}\bigl ({\cal A}\pm {\cal B}\bigr )^{ab}
-l^{[a}\partial^{b]}\bigl ({\cal A}\mp {\cal B}\bigr )_{\mu\nu}-
\nonumber\\
&&
-l_{[\mu}\bigl({\cal A}\mp {\cal B}\bigr )_{\nu]}{}^{\alpha}
\bigl ({\cal A}\pm {\cal B}\bigr)_\alpha{}^{[a}l^{b]}\ .
\end{eqnarray}
The expression for the torsionful curvature satisfies the
following interchange identity:

\begin{equation}\label{eq:interchange}
R_{\mu\nu,\rho\sigma}(\Omega_\pm)= R_{\rho\sigma,\mu\nu}(\Omega_\mp)\
{}.
\end{equation}

We are now ready to investigate the supersymmetry properties of our
Ansatz. We want to find the supersymmetry transformations that leave
the fermions invariant (i.e. equal to zero), that is, all the
non-trivial
supersymmetry transformation parameters $\epsilon$ for which the
r.h.s of eqs. (\ref{eq:susy1})-(\ref{eq:susy3}) vanish (the Killing
spinors). As we
explained in the introduction, we consider only solutions with  at
most one half of
the possible supersymmetries broken. From
eqs.(\ref{eq:susy1})-(\ref{eq:susy3})
we get respectively
\begin{equation}
\phi=\phi(u)\ ,
\hskip 2truecm
[\partial_\mu - {1\over 8}l_\mu
\bigl ({\cal A}+{\cal B}\bigr )_{ij}\gamma^{ij}]\epsilon=0 \ ,
\hskip 2truecm \gamma^u \epsilon =0 \ .
\end{equation}
One half of the supersymmetries are always broken. Also,
$\epsilon$ is a function of $u$ only, and this implies that  $\bigl
({\cal A}+
{\cal B}\bigr)_{ij} \equiv f_{ij}(u)$ must be a function of $u$ only
and

\begin{equation}
\epsilon(u) = e^{1/8
\{\int^u \bigl (f_{ij}(u) \gamma^{ij}\bigr)\}}
\epsilon_0 \ ,
\end{equation}
where $\epsilon_0$ is a constant spinor which satisfies the algebraic
constraint
$\gamma^u \epsilon_{0} =0$.
The integrability condition for the existence of Killing spinors is,
of course, satisfied:

\begin{equation}\label{int}
R_{\mu\nu}{}^{ab}\bigl (\Omega_+)\, \gamma_{ab} \, \epsilon  =
-l_{[\mu}\partial_{\nu]}\bigl ({\cal A} + {\cal B}\bigr )^{ij}
\gamma_{ij} \, \epsilon =0\ .
\end{equation}
Finally, substitution of our Ansatz into the equations of motion,
eqs. (\ref{eq:eqs1})-(\ref{eq:eqs4}), leads to
\begin{eqnarray}
&&\triangle K + 2\partial^i A_i' +{1\over 2} {\cal A}^{ij}{\cal
A}_{ij}
-{1\over 2}{\cal B}^{ij}{\cal B}_{ij} + 4 \phi'' - 2\beta {\rm tr}
\partial^iV
\partial_iV  =0 \ ,\\
&&\triangle V = \partial_i {\cal A}^{ij} = \partial_i {\cal B}^{ij} =
0 \ ,
\end{eqnarray}
where $K=2A_u$, and $\triangle=\partial_{i}\partial_{i}$ is the flat
space
Laplacian.

\bigskip
\section{String Corrections}

Until now we have treated the constants $\alpha$ and $\beta$ as
independent
parameters.
 The field
equations could be solved and half of the supersymmetries preserved
without
relating these parameters to each other. It is well known, however,
that in
string
theory
both $\alpha$ and $\beta$ are related to the Regge slope parameter
$\alpha^{\prime}$ (inverse string tension)  as follows:
\begin{equation}
  \beta = \frac{1}{15} \alpha^{\prime} \ , \qquad \alpha =2
\alpha^{\prime}\ .
\end{equation}
Thus, in this section, we will treat  the classical part of the
Yang-Mills
action
$F^2$ as coming  from string corrections. Here, our starting point
will be the
zero
slope limit ($\alpha^{\prime} \rightarrow 0$) of the bosonic part of
the
effective
action . This is given by
\begin{equation}\label{S0}
S^{(0)}= \frac{1}{2} \int d^{10}x\ \sqrt{-g}e^{-2\phi}\biggl
(-R+4(\partial\phi)^2- {1\over 3}(H^{(0)})^{2} \biggr)\ ,
\end{equation}
where
\begin{equation}\label{H0}
H^{(0)}_{\mu\nu\rho}=\partial_{[\mu}B_{\nu\rho]}\ .
\end{equation}
The superscript in
$H^{(0)}$ indicates that no $O(\alpha^{\prime})$ corrections of any
kind are
 present.
Below we will see that the definition of the axion curvature $H$
contains
an infinite series of higher order
string corrections.

In this section we are going to consider one particularly simple
choice of the
functions
$A_{i}$ and $B_{i}$ of our Ansatz that solves the above theory:
 \begin{equation}\label{choice}
A_{i}=-B_{i} \ , \qquad  f_{ij}(u)=0 \ .
\end{equation}
 The
Killing spinor in our coordinates is a constant spinor satisfying
$\gamma^u
\epsilon_0=0$ \footnote{ Another
example of  an interesting supersymmetric solution with a constant
Killing spinor
in cartesian coordinates is the
purely magnetic extreme dilaton black-hole solution in $d=4$ in
stringy metric
\cite{Ka1}}.  In a recent paper
by Tseytlin \cite{Ts1} the solution in which the vector function in
the metric
is related to the one in the axion was
mentioned as the most natural one from the point of view of the sigma
model
equations.
 It is therefore very interesting to investigate the string
corrections
to the following class of supersymmetric plane waves.
For convenience we rewrite here the different fields and the
equations they
must
satisfy in order to be solutions of the theory (\ref{S0}):
\begin{equation}\label{anzlim}
g_{\mu\nu}  =  \eta_{\mu\nu}+2l_{(\mu}A_{\nu)}\ ,\hspace{2cm}
B_{\mu\nu}  =   3l_{[\mu}A_{\nu]}\ ,\hspace{2cm} V_{\mu}=Vl_{\mu}\ ,
\end{equation}
where $l^\mu A_\mu = A_{v}=0$.  All fields are independent of the $v$
coordinate and, furthermore, $\phi$ is a function of $u$ only. We
have already
included the Yang-Mills fields in the Ansatz (\ref{anzlim}) but we
will not
consider
them until we discuss first order corrections. The equations that
have to be
satisfied are \begin{equation}\label{lapl}
\triangle K + 2\partial^{i} A_{i}^{\prime} + 4 \phi^{\prime\prime}=
\triangle V = \triangle{\cal A}^{ij} = 0\ .
\end{equation}

We get the following derived quantities
\begin{eqnarray}
 H^{(0)}_{\mu\nu\rho} & = & -\frac{3}{2}l_{[\mu}{\cal
                            A}_{\nu\rho]} \ ,\label{Hzero} \\
\Omega_{+\mu}^{(0)}{}^{ab} & = &
       \omega_{\mu}{}^{ab}(e)+H_{\mu}^{(0)}{}^{ab}=-2l^{[a}{\cal
                                  A}^{b]}{}_{\mu}\ ,
\label{Omegazeroplus} \\
\Omega_{-\mu}^{(0)}{}^{ab} & = & \omega_{\mu}{}^{ab}(e)-
                                  H_{\mu}^{(0)}{}^{ab}=+{\cal
                                  A}^{ab}l_{\mu}\ ,
\label{Omegazerominus}\\
R_{\mu\nu}^{(0)}{}^{ab}(\Omega_{+}) & = & -2l^{[a}\partial^{b]}{\cal
A}_{\mu\nu}\ , \\
R_{\mu\nu}^{(0)}{}^{ab}(\Omega_{-}) & = & -2l_{[\mu}\partial_{\nu]}
{\cal A}^{ab}\ .\label{Rzerominus}
\end{eqnarray}
Observe that the expressions above are true even though we only have
related
the
transverse components of $A$ and $B$ in eq. (\ref{choice}), and we
have
said nothing about $A_{u}$ and $B_{u}$. The reason is that $B_{u}$
enters in
these
equations only through $H^{(0)}$, and, actually, it does not
contribute to any
of the
components $H^{(0)}_{\mu\nu\lambda}$ (it is "pure gauge"). This fact
offers the
possibility of choosing $B_{u}$ so that ${\cal A}_{ab}=-{\cal
B}_{ab}$.

 Now let us consider the $\alpha^{\prime}$ corrections. It
is well known that  in the calculation of the string effective action
one has
to add
to $S^{(0)}$ the Lorentz and Yang-Mills Chern-Simons terms which play
a
crucial role in the Green-Schwarz anomaly cancellation mechanism.
These terms
break supersymmetry. To restore supersymmetry order by order in
$\alpha^{\prime}$, one has to add to $S^{(0)}$  an infinite series
of higher
order
terms in $\alpha^{\prime}$ \footnote{The supersymmetrization can be
achieved
either by the Noether method \cite{Ro1} or by superspace methods (for
a
recent review of the latter method, see \cite{Fr1} and references
therein).}.
  By the
procedure of adding terms to restore supersymmetry, the
$O(\alpha^{\prime 3})$
effective action was obtained in \cite{Be1}
 \begin{eqnarray}\label{eric}
S & = & {1\over 2}\int d^{10}x\ \sqrt {-g}e^{-2\phi}\biggl (
-R+4(\partial\phi)^2- {1\over 3}H^2+ \nonumber \\ &&+{1\over 2}T +
2\,
\alpha^{\prime} T^{\mu\nu} T_{\mu\nu}
 +6 \, \alpha^{\prime} T^{\mu\nu\lambda\rho}
T_{\mu\nu\lambda\rho}+ O(\alpha^{\prime 4})\biggr )\ ,
\end{eqnarray}
where the antisymmetric tensor
$T_{\mu\nu\lambda\rho}$, the symmetric tensor $T_{\mu\nu}$ and
the scalar $T$ are given by
\begin{eqnarray}
T_{\mu\nu\lambda\rho} &=&2  \alpha^{\prime} \biggl ( \,
R_{[\mu\nu}{}^{ab}\bigl
(\Omega_-)
                          \, R_{\lambda\rho]}{}^{ab}\bigl (\Omega_-)
+
                           \frac{1}{30}\, {\rm tr} F_{[\mu\nu}
                           F_{\lambda\rho]} \biggr )\label{eq:t1} \
,\\
T_{\mu\nu} &= &2  \alpha^{\prime} \, \biggl (
R_{\mu\lambda}{}^{ab}\bigl
                         (\Omega_-)
                R^{\lambda } {}_{ \nu} {}^{  ab}(\Omega_-)
                + \frac{1}{30} \, {\rm tr} F_{\mu  \lambda}
                F^{\lambda}{}_{\nu} \biggr )\label{eq:t2} \ ,\\
T & = & T_\mu{}^\mu \ \ .
   \label{eq:t3}
\end{eqnarray}
In the above expression there are explicit {\it and} implicit
$\alpha^{\prime}$
corrections. The explicit corrections always appear via $T$-tensors,
and  they
are
essentially the $\alpha^{\prime}$ factors in front of eqs.
(\ref{eq:t1})-(\ref{eq:t3}).
The implicit $\alpha^{\prime}$ corrections always appear via the
torsion $H$
which is
defined by the following iterative procedure: At
the lowest order $H$ is just $H^{(0)}$, defined in eq. (\ref{H0}) and
it is
given by eq.
(\ref{Hzero}) for our solutions. With $H^{(0)}$ we calculate the
lowest
order $\Omega_{\pm}=\Omega_{\pm}^{(0)}$ by using its definition eq.
(\ref{eq:torsion}). $\Omega_{+}$ and $\Omega_{-}$ are given in eqs.
 (\ref{Omegazeroplus}) and
(\ref{Omegazerominus}) for our solutions.  At first order in
$\alpha^{\prime}$,
$H=H^{(1)}$ is $H^{(0)}$ corrected  by the Yang-Mills Chern-Simons
term and the
Lorentz Chern-Simons term corresponding to the zero-order
$\Omega_{-}=\Omega_{-}^{(0)}$:
\begin{eqnarray}
H^{(1)}_{\mu\nu\rho} & = &\partial_{[\mu}B_{\nu\rho]}-\frac{1}{5}
\alpha^{\prime}{\rm tr}
\biggl \{ V_{[\mu}\partial_\nu V_{\rho]}-
{1\over 3}V_{[\mu}V_\nu V_{\rho]}\biggr \}-\nonumber \\
&&-6 \, \alpha^{\prime}  \biggl
 \{ \Omega^{(0)}_{[\mu-}{}^{ab}\partial_\nu
\Omega^{(0)}_{\rho]-}{}^{ab}-
{1\over 3}\Omega^{(0)}_{[\mu-}{}^{ab}\Omega^{(0)}_{\nu-}{}^{ac}
\Omega^{(0)}_{\rho-]}{}^{cb}\biggr \}\ ,\label{susy torsion}
\end{eqnarray}
With $H^{(1)}$ one would get $\Omega^{(1)}$ using again its
definition eq.
(\ref{eq:torsion}) and $H^{(2)}$ would be given by the above
expression but
with
$\Omega^{(0)}$ replaced by $\Omega^{(1)}$. Iterating this procedure
one gets
the
all-order expression $H$ for the torsion which involves the promised
infinite
series of
corrections. Of course, by this procedure, any tensor containing the
torsion
(or the torsionful spin connection) receives an infinite number of
implicit
string
corrections. This applies to the $T$-tensors as well.

One may verify that the lowest order Lorentz Chern-Simons term and
the
Yang-Mills
Chern-Simons term vanish identically for these solutions.
This means that $H^{(1)}=H^{(0)}$, and this in turn
implies that $\Omega^{(1)}=\Omega^{(0)}$ etc. The
conclusion is that  all the implicit string correction vanish for
this class
 of solutions and the expressions in eqs.
(\ref{Hzero})-(\ref{Rzerominus}) are
exact to all orders.

Next we have to study the $T$-tensors for these solutions. One may
easily
establish that the all-order expressions for them are
\begin{eqnarray}
T_{\mu\nu\lambda\rho} & = & T  =  0 \ ,        \\
T_{\mu\nu} & = & -2 \alpha^{\prime} \,l_{\mu} l_{\nu} \biggl\{
   (\partial_{k}{\cal A}^{ij})^2 -\frac{1}{30} \,
(\partial_{k} V^{IJ} )^{2} \biggr\} \ . \label{T}
\end{eqnarray}
Hence, the squares of all these tensors, which we
need to know to calculate the corrections to the on-shell action,
vanish for
these solutions:
\begin{equation}
(T_{\mu\nu\lambda\rho})^2 = (T_{\mu\nu} )^2 = T^2 = 0\ .
\end{equation}
Thus the lowest order effective on-shell string action gets no
corrections.
This is in
agreement with the general non-renormalization theorem for the
on shell action on bosonic solutions with unbroken supersymmetries,
which was
presented in \cite{Ka1}.

Finding the corrections to the equations of motion is more
complicated. To
study them we
first have to vary the action (\ref{eric}) over all the fields
present in the
theory,
and  only then substitute the solutions in the corrected equations.
It is
convenient to
start by studying the linear corrections separately. The equations of
motion
corrected
up to first order come from the action
\begin{equation}
S^{(1)}={1\over 2}\int d^{10}x\ \sqrt
{-g}e^{-2\phi}\biggl (-R+4(\partial\phi)^{2}
-{1\over 3}H^{2} +\frac{1}{2}T \biggr )\ .
 \end{equation}
Note that the notation
$S^{(1)}$ implies by definition that all terms of order
$\alpha^{\prime 2}$ and
higher are
neglected. This applies in
particular to the implicit $\alpha^{\prime}$ dependence present in
$H$ and $T$.

The corrections linear in $\alpha^{\prime}$ to the lowest order
equations of
motion are
derived from the variation  $\delta (S^{(1)}-S^{(0)})\equiv \delta
\Delta S$.
It is
convenient to perform this variation with respect to the explicit
dependence of
the action on
$g_{\mu\nu}$, $V_{\mu}$, $\phi$ and $B_{\mu\nu}$, and then with
respect to
the implicit dependence on these fields through the torsionful spin
connection
$\Omega_{-}$, that is
 \begin{eqnarray}
\delta \Delta S & = &
\frac{\delta\Delta S}{\delta g_{\mu\nu}} \delta g_{\mu\nu}
+\frac{\delta\Delta S}{\delta B_{\mu\nu}}\delta B_{\mu\nu}
+\frac{\delta\Delta S}{\delta\phi}\delta\phi+ \nonumber \\
&&+\frac{\delta\Delta S}{\delta V_{\mu}}\delta V_{\mu}
+\frac{\delta\Delta S}{\delta \Omega_{-\mu}{}^{ab}}
\delta\Omega_{-\mu}{}^{ab}\ ,\label{DS}
\end{eqnarray}
where
\begin{equation}
\delta\Omega_{-\mu}{}^{ab}=
\frac{\delta\Omega_{-\mu}{}^{ab}}{\delta g_{\mu\nu}} \delta
g_{\mu\nu}
+\frac{\delta\Omega_{-\mu}{}^{ab}}{\delta B_{\mu\nu}}\delta
B_{\mu\nu}
+\frac{\delta\Omega_{-\mu}{}^{ab}}{\delta V_{\mu}}\delta V_{\mu}\ .
\end{equation}
The explicit variations are
\begin{eqnarray}
\frac{\delta\Delta S}{\delta g_{\mu\nu}} & = &
-\frac{1}{4}\sqrt{-g}e^{-2\phi}(T^{\mu\nu}-g^{\mu\nu}T)\ ,
\\
\frac{\delta\Delta S}{\delta \phi} & = &
-\frac{1}{2}\sqrt{-g}e^{-2\phi}T\ ,  \\
\frac{\delta\Delta S}{\delta B_{\mu\nu}}
& = & \frac{1}{3}\,
\partial_{\lambda}[\sqrt{-g}e^{-2\phi}
(H^{(1)}{}^{\lambda\mu\nu}
-H^{(0)}{}^{\lambda\mu\nu})] \ ,     \\
\frac{\delta\Delta S}{\delta V_{\mu}} & =
&\frac{1}{15}\alpha^{\prime}\biggl\{
                                        \partial_\lambda
                                     \bigl ( \sqrt{-g}e^{-2\phi}
F^{\lambda\mu}\bigr)
+\sqrt{-g}e^{-2\phi}[V_\lambda
,F^{\lambda \mu}]+
                                         \nonumber \\
                                 &&+  \sqrt{-g}e^{-2\phi}
H^{(0)\mu}{}_{\lambda\rho}
                                 F^{\lambda\rho}-V_{\rho}
                               \partial_{\lambda}(\sqrt{-g}e^{-2\phi}
                                 H^{(0)}{}^{\lambda\mu\rho})\biggr
\}\ .
\end{eqnarray}
Clearly, the r.h.s. of the last three equations vanish for these
solutions,
while
the first equation reduces to
\begin{equation}
\frac{\delta\Delta S}{\delta g_{\mu\nu}}= \frac{1}{2}\alpha^{\prime}
e^{-2\phi}l^{\mu}
l^{\nu} \biggl\{
   (\partial_{k}{\cal A}^{ij})^2 -\frac{1}{30}
(\partial_{k} V^{IJ} )^{2} \biggr\}\ .
\end{equation}

We next consider the implicit variations represented by the last term
in eq.
(\ref{DS}).
In Appendix A we explain a lemma proven in ref. \cite{Be1} that shows
that all
these
implicit variations are proportional to the lowest order equations of
motion of
the
different fields. We therefore conclude that these solutions of the
lowest
order equations of motion are solutions also of the equations of
motion
corrected to
order $\alpha^{\prime}$ if $T_{\mu\nu}=0$. The latter condition can
be
fulfilled by
embedding the torsionful spin connection in an $SO(8)$ subgroup of
the gauge
group and identifying
\begin{equation}\label{embedding}
\partial_{k}{\cal A}^{ij}= \frac{1}{\sqrt{30}}\partial_{k}V^{IJ}\ .
\end{equation}
Here the Yang-Mills indices $IJ$ refer to the adjoint representation
of
$SO(8)$.
Notice that $SO(8)$ is contained both in $SO(32)/Z_{2}$ and in
$E(8)\times
E(8)$. In view of their very special properties, we will call the
specific
class of
solutions that satisfy (\ref{embedding})  {\it supersymmetric
string waves} (SSW) . This concludes the discussion of the
corrections
linear in $\alpha^{\prime}$.

Now let us consider the higher order corrections. The general
structure of the
bosonic part of the effective action, which can be obtained by the
procedure
outlined before based upon the restoration of supersymmetry, has been
conjectured in ref. \cite{Be1} to be of the form
\begin{equation}
{\cal L}_{eff}= \sum_{n=0,1,2,\dots}\alpha^{\prime n}\bigl( a_{n}
R \, T^{n}+b_{n}T^{n+1}\bigr)\ ,
\end{equation}
where $R$ is the Riemann curvature tensor and
$T\equiv\alpha^{\prime}(R^{2}+F^{2})$.
This has been proven up to order $\alpha^{\prime 4}$ in ref.
\cite{Be1}. The
terms
proportional to $a_{0}$ and $b_{0}$ have just been discussed. It is
known that
there are
no terms proportional to $a_{1}$. The $b_{1}$ terms are given by
\begin{equation}
\alpha^{\prime}\int d^{10}x\ \sqrt{-g}e^{-2\phi}\
(T^{\mu\nu}T_{\mu\nu}+3T^{\mu\nu\lambda\rho}T_{\mu\nu\lambda\rho})\ .
\end{equation}
In the variations of the higher order terms, one should again
distinguish
between
the variations with respect to the explicit dependence on the fields
$g_{\mu\nu},\phi$ and $V_{\mu}$ and the variations with respect to
the implicit
dependence of these terms on the previous fields and $B_{\mu\nu}$
through the
torsionful spin-connection $\Omega_{-}$. It is easy to see that the
variations
of
the first kind do not contribute to the field equations for the SSW.
The same
applies to the second kind of variations. As an example, we show how
this works
for the $T^{2}$ terms written above.

Varying with respect to the spin connection and using the fact that
$T^{\mu\nu\rho\lambda}=0$ for the SSW, one finds that the only
remaining terms
are
\begin{equation}
T^{\mu\nu}R_{\mu}{}^{\lambda}{}_{ab}(\Omega_{-})[D_{\lambda}(\Omega_{-
})\delta
\Omega_{-\nu}{}^{ab}- D_{\nu}(\Omega_{-})\delta
\Omega_{-\lambda}{}^{ab}] \ .
\end{equation}
For the SSW solution $T^{\mu\nu}\sim l^{\mu}l^{\nu}$ , and
\begin{equation}\label{kill}
l^{\mu}R_{\mu\nu}{}^{ab}(\Omega_{-})=l_{a}R_{\mu\nu}{}^{ab}(\Omega_{-}
)=0\  .
\end{equation}
Hence, the remaining terms also vanish.

Finally, we consider the corrections to the
supersymmetry transformation laws and the Killing spinors.
In ref. \cite{Be1} it was shown that the corrections to order
$\alpha^{\prime 4}$
are proportional to $T$-tensors and at most one torsionful curvature
tensor.
Therefore,
these corrections vanish for SSW that satisfy the embedding equation
(\ref{embedding}).
Actually, this condition is not necessary in order to establish the
vanishing
of the
corrections to the Killing spinors. The reason is that without the
condition
(\ref{embedding}), the only non-vanishing $T$-tensors are
$T_{\mu\nu}\sim
l_{\mu}l_{\nu}$. Since our fermionic fields have at most one free
index, one of
the indices
of $T_{\mu\nu}$ has to be contracted either with a curvature tensor
or with an
antisymmetrized gamma matrix $\gamma^{[\mu_{1}\dots\mu_{n}]}$. In the
first
case the
correction vanishes due to eq. (\ref{kill}), and in the second case
due to the
fact that
$l_{\mu}\gamma^{\mu}\epsilon=0$. Our Killing spinors, which were
obtained in
the  zero order approximation, satisfy the corrected equations as
well.

As we have discussed before, we have only considered the terms that
have to be
added
to the effective action in order to supersymmetrize the Lorentz and
Yang-Mills
Chern-Simons term.  It is well known, however, that other terms occur
in the
superstring
effective action that are not related by supersymmetry to the Lorentz
and
Yang-Mills  Chern-Simons forms,
for instance terms of the form $\zeta (3)R^{4}$ \cite{Gri1},
\cite{Gro1}\footnote{The
supersymmetrization of the most general $R^{4}$ terms has recently
been
considered
in ref. \cite{Roo1}.}. Nevertheless, all the terms found so far are
at least
quartic in the
Riemann or Yang-Mills curvature tensors or in the torsion tensor and,
therefore, by
simple null vector counting, they are harmless.

Our final conclusion is that the on-shell action, the  fields that
solve
the lowest order equations of motion and the Killing spinors for the
SSW
solutions do not
receive any higher order string corrections.

\section{Summary and Conclusion}

In this paper we have found a quite general class of plane wave type
solutions
of superstring theory, with one half of the space-time
supersymmetries unbroken
and whose 8 Killing spinors are constant. They are solutions to the
field
equations of the
zero slope limit of the  superstring effective action and do not
receive any higher order string corrections.
The metric, the axion and the gauge fields of these solutions are all
described
in terms of {\it one vector function } of  the light-cone coordinate
$u$ and of
the transverse coordinates $x^i$,
\begin{equation}
A_\mu (u, x^i)=  \{ A_{u}\equiv \frac{1}{2}K(u, x^i), \,  A_{v}=0, \,
A_{i}(u,
x^i) \}\ , \quad
i=1,...,8 \ .\label{vec}
 \end{equation}
They are given by
\begin{equation}
g_{\mu\nu}  =  \eta_{\mu\nu}+2l_{(\mu}A_{\nu)}\ ,\hspace{2cm}
H_{\mu\nu\lambda}  =   3\partial_{[\mu} l_{\nu}A_{\lambda]}\
,\hspace{2cm}
F_{\mu\nu}^{IJ}= 2 \sqrt{30}\, \partial_{[\mu}{\cal A}^{ij} l_{\nu]}\
.\label{SSW}
\end{equation}
where ${\cal A}^{ij} = \partial^{[i} A^{j]}$ .
The dilaton  $\phi(u)$  is a function of $u$ only. The equations that
$A_{u} = \frac{1}{2} K(u, x^i)$ and $ A_{i}(u, x^i)$
have to satisfy are
\begin{eqnarray}
&&\triangle K + 2\partial^i A_i' + 4 \phi''  =0 \ , \\
&&\triangle{\cal A}^{ij} = 0\ .
\label{la}\end{eqnarray}
These solutions are based upon a special conspiracy between the
geometry, the
axion
field and  the gauge field. In the non-supersymmetric plane waves
considered in
\cite{Ho1}, \cite{Ts1} there are no relations between
the metric, the axion and the gauge fields, besides
those coming from the equations of motion. In the zero slope limit
the plane
waves are supersymmetric if the geometry and the axion field
are related by $\partial_i{\cal A}^{\mu \nu} = - \partial_i{\cal
B}^{\mu
\nu}$. The requirement of absence of
$\alpha^{\prime}$ corrections to the zero slope supersymmetric
solutions puts an even stronger constraint on the solutions: it
forces us to
embed the torsionful spin connection in the gauge group and the gauge
field is
also now expressed in terms of the same vector field $A_\mu (u,
x^i)$.
This is possible due to the fact that both ${\cal A}^{ij}$ and
$V^{IJ}$ satisfy
the same
(harmonic) equations of motion.

Our SSW solutions extend  the supersymmetric
exact plane wave solutions of string theory studied before by
G\"{u}ven
\cite{Gu1} in the
following sense. In  G\"{u}ven's solutions, the function  $K(u, x^i)$
is
quadratic in the $x^{i}$,
the non-diagonal functions in the metric  $A_{i}(u, x^i) dx^i du $
are absent,
and the axion $H_{uij}$ is
a function of $u$ only, i.e. the dependence on $x^i$ in $B_{ui}$ is
linear in
$x^i$. There are no
relations between the geometry, axion and gauge fields, besides those
coming
from the equations of motion. Due to the restricted  dependence of
these
solutions
on the transverse dimensions, it was not necessary to relate the
axion,
graviton and gauge
fields. Also, there was no need to know the explicit form of the
string
corrections
to prove that
they vanish. The existence of  a covariantly constant null vector
along with
the
simple dependence  on
$x^i$ was sufficient to establish the non-renormalization theorem.

The SSW, which we have studied, have unconstrained dependence on the
transverse
coordinates $x^i$. The price to pay for making the corrections vanish
is the
existence of
specific relations between different fields in the solution.  This
means, for
example, that
generalizations of the Aichelburg-Sexl geometry, where functions like
$\ln (x_i
x^i)$  are present, or solutions where any other complicated
dependence on
$x^i$
appears in a metric of the general form (\ref{eq:metric}),  can be
investigated
in the
framework of this SSW. In this paper we have shown that such metrics
with the
inclusion of
appropriate dilaton, axion and gauge fields also belong to the class
of
perturbatively exact
stringy supersymmetric solutions.

It is interesting to compare our SSW solutions with the
supersymmetric string solitons (SSS) studied in \cite{Da1}. The
common
features are the
following. In both cases it is crucial that
the supersymmetry transformations depend on $\Omega_{+\mu}{}^{ab}$
torsionful spin connections, and  that the
$\alpha^{\prime}$  string corrections depend on
$\Omega_{-\mu}{}^{ab}$.
Both for SSW and SSS the embedding of the spin connection into the
gauge
group is a necessary condition for the absence of higher order string
corrections.
In addition, the interchange identity between the two types of
curvature
(\ref{eq:interchange}) ensures the consistency of the solutions.
In both cases there exists a conspiracy  between the metric and the
axion
field.

A difference, however, is that the dilaton plays  a crucial role in
the
construction of the SSS solution but
does not play any particular role in the  SSW solution.
Furthermore, in the SSS case, the existence of supersymmetry relies
on the
four-dimensional space-time self-duality and the unbroken
supersymmetries are
given by  chiral spinors with $\epsilon_+ = 0$.   In  the SSW case,
the
preserved
supersymmetries depend on the existence of the covariantly constant
null vector
$l_{\mu}$ and the
Killing spinors are  constant spinors $\epsilon$ such that
$l_{\mu}\gamma^{\mu}\epsilon = 0$.

Finally, it would be interesting to investigate special examples of
our SSW
solutions and apply  sigma
model duality transformations \cite{Bu} to study how the
supersymmetry and
non-renormalization properties behave under them. In particular, it
has been
established before \cite{H} that the metric outside an extremal
fundamental
string is the
dual of a plane fronted wave metric with a dilaton and $
A_u$-function in the
metric, that
describes a string boosted to the speed of light. These plane waves
have
vanishing axion
field and vanishing function $A_i$ in the metric and no Yang-Mills
field.
They satisfy our
definition of supersymmetric string waves, given in eqs. (\ref{SSW}).
  They
are the
trivial case in which the conspiracy between metric, axion and gauge
field is
achieved by the choice $A_i= 0$. It would be interesting to
investigate
whether the more general supersymmetric string waves studied in
our paper generate new interesting geometries via sigma model
duality.

On a more fundamental level, one may try to develop a new approach to
string quantum gravity starting with supersymmetric string waves.
The standard
quantum field theory approach is based on the flat-space ``plane
waves"
$\sim e^{ik \cdot x}$, which are solutions of the linearized Einstein
equations.
 The special class of gravitational waves  investigated in this
paper, which are
solutions of the non-linear Einstein equations, and even more, of the
 non-linear Einstein equations with all $\alpha^{\prime}$-corrections
taken into account, may serve as the basis for a new expansion of
the path integral for quantum
gravity.

\section*{Acknowledgements}

The authors wish to thank G. Horowitz for bringing to our
attention the issue of the non-renormalization theorem for
gravitational waves.
We are grateful to A. Linde and A. Peet for useful discussions.

The work of E. B., R.K. and T.O. was partially supported by a NATO
Collaborative Research Grant. The work of E. B. has been made
possible by a
fellowship of the Royal Netherlands Academy of Arts and Sciences
(KNAW). E. B.
would like to thank the Physics Department of Stanford University for
its
hospitality. The work  of  R.K.  was supported in part by NSF grant
PHY-8612280 and Radway Fellowship in
the School of Humanities and Sciences at  Stanford University.
 The work of T.O.
was supported by a Spanish Government M.E.C. postdoctoral grant.

\newpage

\appendix

\section{Lemma}

In this Appendix we give the following Lemma which is used in the
text when
discussing the
string corrections. A proof of (the supersymmetric version of) this
Lemma can
be found in
\cite{Be1}. \bigskip

\indent {\it Lemma.} \hskip .5truecm For arbitrary transformations
 $\delta \Omega_{\mu -}^{ab}$
the variation of the action

\begin{equation}
S={1\over 2}\int d^{10}x\ \sqrt {-g}e^{-2\phi}\biggl (
-R+4(\partial\phi)^2-
{1\over 3}H^2 + {1\over 2}T \biggr )
\end{equation}

\noindent is given by
\begin{equation}
\delta S =2  \alpha^{\prime}\int d^{10}x \ \bigl (\delta_1 {\cal L}
 +\delta _2 {\cal L}\bigr )
\end{equation}
with
\begin{eqnarray}
\delta_1 {\cal L} &=& 6 {\cal B}^{\mu\nu}\Omega_{\mu -}^{ab}\delta
\Omega_{\nu -}^{ab} +2\bigl ( 4 \tilde{{\cal E}}_{\rho a}- e_{\rho
a}\tilde{\Phi}
+6
\tilde{{\cal B}}_{\rho a}\bigr )
\bigl [(\sqrt {-g})^{-1}e^{2\phi} D_\lambda(\Omega_+)(\sqrt
{-g}e^{-2\phi}
e^{\nu a}\delta \Omega_{\nu -}^{\lambda\rho})\bigr ]\  ,\\
&&\nonumber\\
\delta_2 {\cal L} &=& 6  D_\mu(\Omega_-)(\sqrt
{-g}e^{-2\phi}T^{\mu\nu ab})
\delta\Omega_{\nu -}^{ab}
+6 \sqrt {-g}e^{-2\phi}H^{\nu\lambda\rho}T_{\lambda\rho
ab}\delta\Omega_{\nu
-}^{ab}\ , \end{eqnarray}

\noindent where $T_{\mu\nu\lambda\rho}$ is defined in (\ref{eq:t1})
and
 $\tilde{\Phi},
\tilde{{\cal E}}_{\mu\nu},
$ and $\tilde{{\cal B}}_{\mu\nu}$ are given by the expression of
$\Phi, {\cal
E}_{\mu\nu},
$ and ${\cal B}_{\mu\nu}$, the lowest order equations of motion
corresponding to the fields $\phi$, $B_{\mu\nu}$ and $g_{\mu\nu}$
respectively
but where $H^{(0)}$
is replaced by $H$.
These lowest order equations are given by
\begin{eqnarray}
\Phi \equiv {\delta S^{(0)}\over \delta\phi} & = &
\sqrt{-g}e^{-2\phi}[R-4\nabla_\mu\partial^\mu \phi+4(\partial\phi)^2
+
 {1\over 3}(H^{(0)})^2] \ ,           \\
{\cal E}_{\mu\nu} \equiv {\delta S^{(0)}\over \delta g^{\mu\nu} } & =
&
-  {1\over
2}\sqrt{-g}e^{-2\phi}(R_{\mu\nu}-2\nabla_\mu\partial_\nu\phi +
 H^{(0)}_{\mu\lambda\rho}H^{(0)}_\nu{}^{\lambda\rho}
- g_{\mu\nu}\Phi)\ ,
\\
{\cal B}^{\mu\nu} \equiv {\delta S^{(0)}\over \delta B_{\mu\nu}} & =
&
{1\over 3} \partial_\lambda [ \sqrt
{-g}e^{-2\phi}(H^{(0)})^{\lambda\mu\nu}]\ .
 \end{eqnarray}

Note that the combination
\begin{equation}
-2 \tilde{{\cal E}}_{\mu\nu} -3 \tilde{{\cal B}}_{\mu\nu} +
{1\over 2}g_{\mu\nu}\tilde{\Phi} =
 \sqrt{-g}e^{-2\phi}\biggl (
R_{\mu\nu}(\Omega_+)-2 \nabla_\mu(\Gamma_+)\partial_\nu\Phi\biggr )
\end{equation}
is precisely
that in which the $H$-dependence can be absorbed into a spin
connection
with torsion $\Omega_{+}$.

\newpage

\section {Notation and conventions}

We use the notation and conventions of \cite{Be1}. However, in
order to conform as much as possible with the recent
literature \cite{Da1} we have made the following redefinitions with
respect to
\cite{Be1}: $\phi\rightarrow e^{2\phi/3}, B_{\mu\nu}\rightarrow
{1\over 3}\sqrt 2 B_{\mu\nu},
A_\mu\rightarrow V_\mu$ and $\lambda\rightarrow -{1\over 4}\sqrt 2
\lambda$.
We use a metric with mostly minus signature\footnote{
Note that in \cite{Be1} the Pauli metric is used. Here we have
converted the results of \cite{Be1} to correspond to a mostly minus
metric.}. Our conventions for the
Riemann tensor are

\begin{equation}
R_{\mu\nu\rho}{}^\sigma=
\partial_\mu
\bigl\{ \matrix{\sigma\cr\nu\rho} \bigr\}
-\partial_\nu
\bigl\{ \matrix{\sigma\cr\mu\rho} \bigr\}
+
\bigl\{ \matrix{\sigma\cr\mu\lambda} \bigr\}
\bigl\{ \matrix{\lambda\cr\nu\rho} \bigr\}
-
\bigl\{ \matrix{\sigma\cr\nu\lambda} \bigr\}
\bigl\{ \matrix{\lambda\cr\mu\rho} \bigr\} \ ,
\end{equation}
where $\bigl\{ \matrix{\rho\cr\mu\nu} \bigr\}$ are the standard
Christoffel
symbols:

\begin{equation}
\bigl\{ \matrix{\rho\cr\mu\nu} \bigr\}= {1\over 2}
g^{\rho\lambda}\biggl\{
\partial_\mu g_{\nu\lambda}+\partial_\nu
g_{\mu\lambda}-\partial_\lambda
g_{\mu\nu}
\biggr\} \ .
\end{equation}
The Ricci tensor and the curvature scalar are defined by:

\begin{equation}
R_{\mu\nu}= g^{\lambda\rho}R_{\mu\lambda\nu\rho}\ , \hskip 3truecm R=
R_\mu{}^\mu\ .
\end{equation}
The standard general covariant derivative $\nabla_{\mu}$ is
\begin{equation}
\nabla_\mu V_\nu = \partial_\mu V_\nu -
\bigl\{ \matrix{\rho\cr\mu\nu} \bigr\}
V_\rho \ , \hskip 2truecm \nabla_\mu V^\nu = \partial_\mu V^\nu +
\bigl\{ \matrix{\nu\cr\mu\rho} \bigr\}V^\rho \ .
\end{equation}

 The supersymmetry transformation rules
(\ref{eq:susy1})-(\ref{eq:susy3})
involve the
zehnbein fields $e_\mu{}^a$ and their inverses $e^a{}_\mu$. They are
related to
the metric tensor and its inverse via
\begin{equation}
g_{\mu\nu} = e_\mu{}^ae_\nu{}^b\eta_{ab}\ ,\hskip 3truecm g^{\mu\nu}
=
e_a{}^\mu e_b{}^\nu\eta_{ab}\ .
\end{equation}
The spin connection field $\omega_\mu{}^{ab}(e)$ is defined in terms
of
derivatives
of the zehnbein fields as follows:
\begin{equation}
\omega_\mu{}^{ab}(e)= -e^{\nu[a}\bigl ( \partial_\mu
e_\nu{}^{b]}-\partial_\nu
e_\mu{}^{b]}\bigr ) - e^{\rho[a} e^{\sigma b]} \bigl (
\partial_\sigma
e_{c\rho}\bigr) e_\mu{}^c\ ,
\end{equation}
and it is related to the Christoffel symbols by
\begin{equation}
\omega_{\mu}{}^{ab}=-e_{\nu}{}^{a}e^{\rho b}\bigl\{
\matrix{\nu\cr\mu\rho}
\bigr\}+e^{\nu b}\partial_{\mu}e_{\nu}{}^{a}\ .
 \end{equation}
The curvature tensor corresponding to this spin connection field is
defined by
\begin{equation}
R_{\mu\nu}{}^{ab}(\omega)=
2\partial_{[\mu}\omega_{\nu]}{}^{ab}-
2\omega_{[\mu}{}^{ac}
\omega_{\nu]c}{}^b\ , \hskip 2truecm
R(\omega)\equiv
e^\mu{}_ae^\nu{}_bR_{\mu\nu}{}^{ab}(\omega)\ .
\end{equation}
It is related to the Riemann curvature tensor as follows:
\begin{equation}
R_{\mu\nu\rho\sigma} =
R_{\mu\nu}{}^{ab}(\omega)e_{a\rho}e_{b\sigma}\ .
\end{equation}
This relation follows as an integrability condition from the
zehnbein postulate
\begin{equation}
\partial_\mu e_\nu{}^a -
\bigl\{ \matrix{\rho\cr\mu\nu} \bigr\}
e_\rho{}^a - \omega_\mu{}^{ab}e_{\nu b} =0\  .
\end{equation}

To specify our Ansatz for the metric, axion and
dilaton fields, it is convenient to first write the ten-dimensional
coordinates
$x^\mu$ in a light-cone basis:
\begin{equation}
x^\mu = (u,v,x^i),\ i=1,\dots ,8, \ \ \  u={1\over \sqrt
2}(x_0+x_9),\ v
={1\over \sqrt 2}(x_0-x_9)\ .
\end{equation}
In this paper, all the indices are raised and lowered with the full
ten-dimensional metric
$g_{\mu\nu}$. In the case in which the metric is given by eq.
(\ref{eq:metric}), the
relation between upper and lower indices is
\begin{eqnarray}
\xi^{u} & = & \xi_{v}\ \ , \\
\xi^{v} & = & \xi_{u}-(2A_{u}+\sum_{i+1}^{8}A_{i}^{2})\xi_{v}+
\sum_{i=1}^{8}A_{i}\xi_{i}\ \ ,\\
\xi^{i} & = &A_{i}\xi_{v}-\xi_{i}\ \ .
\end{eqnarray}
Note that if $\xi_{v}=0$, the transverse indices $i$ can be raised
and lowered
with the
flat metric.

\pagebreak

\end{document}